%% file: main.tex
\documentclass{article}

\usepackage{arxiv}
\usepackage[sort&compress,square,comma,numbers]{natbib}
\usepackage[utf8]{inputenc} %
\usepackage[T1]{fontenc}    %
\usepackage{hyperref}       %
\usepackage{url}            %
\usepackage{booktabs}       %
\usepackage{graphicx}

\usepackage{doi}

\usepackage{amsmath}
\usepackage{hhline}
\usepackage{cleveref}
\usepackage{xcolor}
\usepackage{multirow}
\usepackage{breqn}
\usepackage[linesnumbered,ruled,vlined]{algorithm2e}
\usepackage{tabu}

\crefname{algocf}{alg.}{algs.}
\Crefname{algocf}{Algorithm}{Algorithms}

\title{MetaMedSeg: Volumetric Meta-learning for Few-Shot Organ Segmentation}
\renewcommand{\shorttitle}{MetaMedSeg: Volumetric Meta-learning for Few-Shot Organ Segmentation}

\newcommand\blfootnote[1]{%
  \begingroup
  \renewcommand\thefootnote{}\footnote{#1}%
  \addtocounter{footnote}{-1}%
  \endgroup
}

\date{} 					%

\author{ Anastasia Makarevich\thanks{The first two authors contributed equally to this work.} \\
	Technical University of Munich\\
	Munich, Germany \\
	\texttt{ana.makarevich@tum.de} %
	\And
	Azade Farshad\footnotemark[1] \\
	Technical University of Munich\\
	Munich, Germany \\
	\texttt{azade.farshad@tum.de} \\
	\And
	Vasileios Belagiannis \\
	Ulm University\\
	Ulm, Germany \\
	\texttt{vasileios.belagiannis@uni-ulm.de} %
	\And
	Nassir Navab \\
	Technical University of Munich\\
	Munich, Germany \\
	\texttt{nassir.navab@tum.de} \\
}

\renewcommand{\headeright}{}
\renewcommand{\undertitle}{}

\hypersetup{
pdftitle={MetaMedSeg: Volumetric Meta-learning for Few-Shot Organ Segmentation},
pdfsubject={ML, CV},
pdfauthor={A.Makarevich, A.Farshad, V.Belagiannis, and N.Navab},
pdfkeywords={meta-learning, few-shot organ segmentation},
}

\begin{document}
\maketitle

\input{Chapters/abstract}
\input{Chapters/intro}
\input{Chapters/method}

\input{Chapters/results}
\input{Chapters/conclusion}

\bibliographystyle{unsrt} %
\bibliography{ref}  %

\end{document}

%% file: Chapters/abstract.tex
\begin{abstract}
The lack of sufficient annotated image data is a common issue in medical image segmentation. For some organs and densities, the annotation may be scarce, leading to poor model training convergence, while other organs have plenty of annotated data. In this work, we present MetaMedSeg, a gradient-based meta-learning algorithm that redefines the meta-learning task for the volumetric medical data with the goal to capture the variety between the slices. We also explore different weighting schemes for gradients aggregation, arguing that different tasks might have different complexity, and hence, contribute differently to the initialization. We propose an importance-aware weighting scheme to train our model. In the experiments, we present an evaluation of the medical decathlon dataset by extracting 2D slices from CT and MRI volumes of different organs and performing semantic segmentation. The results show that our proposed volumetric task definition leads to up to $30\%$ improvement in terms of IoU compared to related baselines. The proposed update rule is also shown to improve the performance for complex scenarios where the data distribution of the target organ is very different from the source organs.\blfootnote{Project page: \url{http://metamedseg.github.io/}}

\end{abstract}

%% file: Chapters/intro.tex
\section{Introduction}
Segmentation of medical images is an effective way to assist medical professionals in their diagnosis. Recent advances in deep learning have made it possible to achieve high accuracy in organ and tumour segmentation~\cite{ronneberger2015u,milletari2016v}. Despite the recent advances in medical image segmentation, standard supervised learning settings usually require a large amount of labelled data. Labelled data can be abundantly available for some organs (e.g., liver), yet it can be really scarce for others. One of the early approaches to overcome this limitation is transfer learning, where a neural network is pertained on a large labelled dataset (source domain) and then fine-tuned on a small amount of labelled data (target domain)~\cite{rohrbach2013transfer}.

Another common approach, that gained a lot of popularity, is few-shot learning which aims to learn from just a few examples.  One recent example of few-shot learning is COVID-19 detection using chest X-rays ~\cite{jadon2021covid}. Few-shot methods can be roughly divided into augmentation-based learning and task-based meta-learning. In this work we focus on the meta-learning approach, which also comes in different flavours: metric learning (e.g., prototypical networks~\cite{snell2017prototypical,wang2019panet,yang2020prototype}), memory-based learning ~\cite{hu2019attention,cao2020few} and gradient-based methods. Few-shot learning for image segmentation has been an active research topic~\cite{tian2020prior,boudiaf2020few,azad2021texture,siam2019amp, gairola2020simpropnet, liu2020crnet, tian2020differentiable}, but there are few works~\cite{mondal2018few,dawoud2020few,ouyang2020self} focusing on few-shot medical image segmentation. %

In this work, we follow the meta-learning approach, relying on one of the recent modifications of the MAML (Model-Agnostic Meta-Learning) algorithm~\cite{finn2017model} - Reptile~\cite{nichol2018first}, which is a simple and yet effective algorithm that provides a wide field for experiments. We address the two main components of gradient-based meta-learning: task definition and gradient aggregation. By gradient aggregation, we mean the update of the weights of the meta-model. The task definition is the concept of creating tasks by sampling pairs of images and their corresponding segmentation maps based on the specified criteria. We propose a volume-based task definition, specifically designed for volumetric data, and introduce a weighting mechanism for aggregation of gradients in each meta-training step, beneficial to non-IID (independent and identically distributed) data. 

The main contribution of this work is the volumetric task definition. We show that sampling data from one volume per task could lead to better optimization of the local models on the specific organ due to more control over the shots variability. In contrast to the standard setting, where tasks are sampled randomly and can as well end up with images from similar parts of the volume (e.g. just first slices of different volumes), we ensure that a certain level of diversity exists between the shots, but at the same time, the diversity of the source set in general is reduced, since each volume is associated with a fixed, limited number of shots (e.g. 15) and other shots never participate in the training, which has shown to have a positive effect on training as shown in ~\cite{setlur2020support}. Our second contribution is the importance-aware weighting scheme. In the classical Reptile setting, gradients of sampled tasks are averaged in each meta-epoch, while in our proposed method, the gradients are weighted based on the importance of each task. The importance is defined as the distance between a local model trained on a given task and the average of all models trained on other tasks. This weighting mechanism has been proposed before in the federated learning framework~\cite{yeganeh2020inverse} for non-IID data. We argue that by giving less weight to tasks with a higher distance from the average model, less chance is given to the outlier data which could help avoid catastrophic forgetting (the tendency of the neural network to forget previously learnt information)~\cite{bertugli2020few}. This could also perform as a regularization to avoid overfitting when tasks are similar and benefit the training when the cross-domain distance is high. 

\begin{figure}[t]
    \centering
    \includegraphics[width=\textwidth]{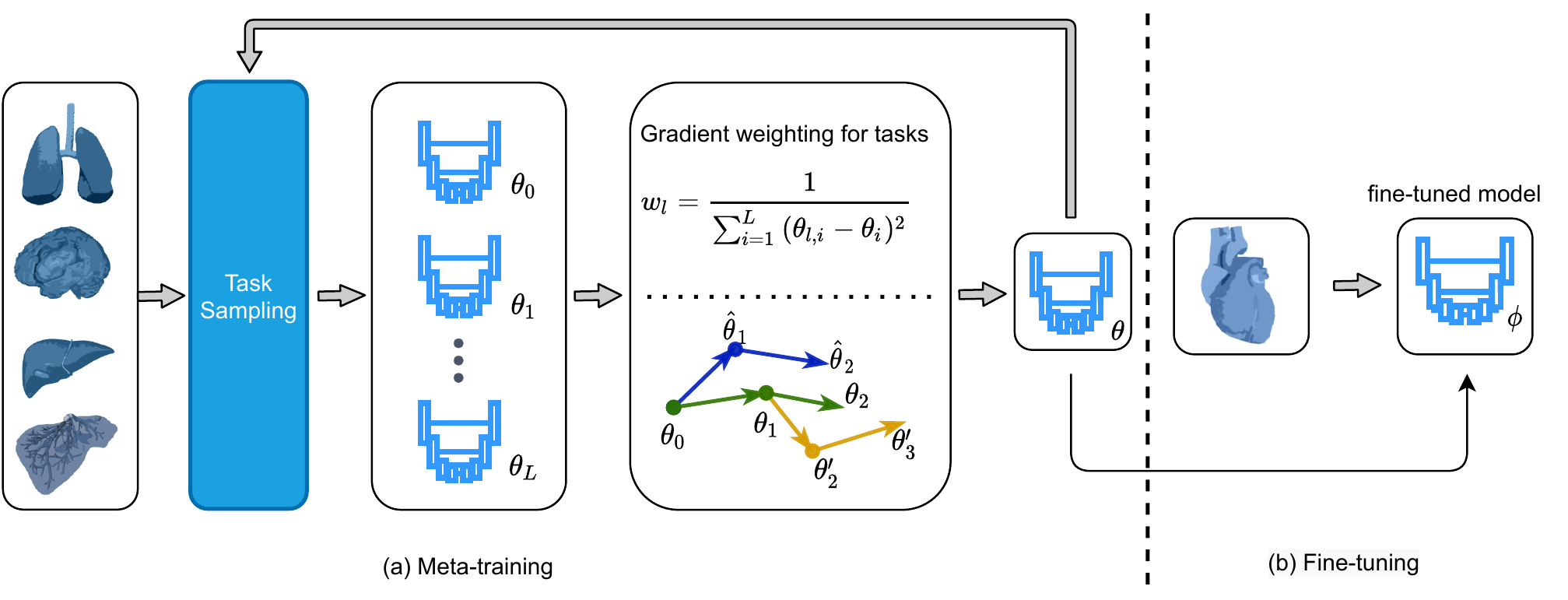}
    \caption{The meta-training step on source organs is shown on the left, where we define the meta-training tasks and weight these tasks based on their importance. $\theta$ is the meta-model parameters and $\phi'$ is the model in the progress of meta-training. The meta-trained model with parameters $\phi$ is then used for the final fine-tuning step on the target organ.}
    \label{fig:teaser}
\end{figure}

Our evaluations show that the proposed volumetric task definition and weighted gradient aggregation improve the segmentation's accuracy. We evaluate our method in two different settings: 1. Few-shot, where the models are fine-tuned on few shots. 2. Full-data, where the model is fine-tuned on all of the data for the target organ. We compare our results with multiple baselines: 1. Supervised learning with random initialization, 2. Supervised learning with transfer learning initialization, 3. Reptile baseline, with and without our proposed volumetric task definition, 4. Few-shot cell segmentation by Dawoud et al.~\cite{dawoud2020few}, which is the closest related work, with and without our proposed volumetric task definition. \Cref{fig:teaser} shows an overview of our method.\\

To summarize, we propose MetaMedSeg, a meta-learning approach for medical image segmentation. The main contributions of this work are as follows: 1. A novel task definition based on data volumes designed for medical scenarios 2. A novel update rule for few-shot learning where the cross-domain distance is high. 3. Significant improvement of segmentation performance compared to standard methods. %

%% file: Chapters/method.tex
\section{Methodology}
Given a dataset $\mathcal{D} = \{\mathcal{S}, \mathcal{T}\}$, we define $\mathcal{S} = \{\mathcal{S}_1 \cdots \mathcal{S}_{n}\}$ as the training set (source domain) and $\mathcal{T} = \{\mathcal{T}_1 \cdots \mathcal{T}_{m}\}$ as the test set (target domain), where $n, m$ are the number of organ datasets for the source and target domains and $\mathcal{T} \cap \mathcal{S} = \emptyset$. Each dataset consists of pairs of images and segmentation masks. The task in our setup is defined then as a subset of k shots sampled from $\mathcal{S}_i$ or $\mathcal{T}_j$.

The learning has two steps: meta-training and fine-tuning. The model parameters from the meta-training step are denoted by $\theta$, while the fine-tuned model parameters are denoted by $\phi$. At each meta-train step, each task learns its own set of parameters denoted as $\theta_l$, starting with meta-model weights $\theta$. The network architecture used in this work is the well-known U-Net~\cite{ronneberger2015u} architecture, which is commonly used for medical image segmentation. The input to the network is a batch of images $\mathcal{I}_b$, the outputs are segmentation maps $y_b$. The ground-truth segmentation maps are denoted by $y'_b$. Instead of batch normalization, we adopt the instance normalization approach ~\cite{bronskill2020tasknorm,jia2019instance} for the meta-learning setting. \Cref{alg:method} shows the base version of the algorithm for our approach. The main components of our work are: 1. \textit{Meta-learning} 2. \textit{Image Segmentation}. We discuss each of them below.

\begin{algorithm}[htbp]
\DontPrintSemicolon
  
  \KwIn{Meta-train datasets $\mathcal{S} = {\mathcal{S}_1, \mathcal{S}_2, ..., \mathcal{S}_{n}}$}
  \KwIn{Meta-test dataset T}
  
  Initialize: $\theta$
  
  \For{meta-epoch = 1,2,...,N}
  {
  
  Sample L datasets from meta-train datasets $\mathcal{S}$
  
  \For{l = 1,2,...,L}
  {
  
  Sample K shots from dataset $\mathcal{S}_l$ using rule $\mathcal{R}$ 
  
  Train base learner (U-Net) to obtain $\theta_l = g(L(\theta, \mathcal{S}_l))$ 

  }
  
  Compute task importance $w_l$
  
   Perform meta-update: $\theta \leftarrow \theta +
  \beta \sum_{l=1}^{L}  {w_l (\theta_l-\theta)}  $

  }
Sample $K'$ shots from meta-test dataset $T$ to generate $T' \in T$

Fine-tune on $T'$

Compute test IoU on $T'' \in T, T' \cap T'' = \emptyset$

\caption{MetaMedSeg for organ segmentation}
\label{alg:method}
\end{algorithm}

\subsection{Meta-learning}
As shown in ~\cref{fig:teaser}, in each round of meta-training, a set of tasks consisting of images and their corresponding segmentation maps are sampled according to the chosen rule and fed to the U-Net model. The learned weights of all of these models are then aggregated based on the specified update rule. The final model is used as initialization for the fine-tuning step. In classical Reptile algorithm~\cite{nichol2018first}, the updates obtained from all tasks are averaged in each meta-epoch. We propose a different strategy for weighting these updates based on the importance of the tasks. The details of task definition, task sampling and update rules will be discussed next.
\subsubsection{Task definition}
A task in meta-learning for segmentation can be defined in different ways. Our initial approach is based on ~\cite{dawoud2020few}. A task is a set of $k$ images and masks belonging to the same dataset. For example, k-shots sampled randomly from all available vessel cancer slices is such a task. This approach is targeted mainly at 2D data, and although we are also working with 2D slices, there is additional information that could be used. Our data is not just a set of images but a set of volumes (3D tensors that can be sliced along chosen direction to produce sets of 2D images), so we propose a volume-based task definition. 
We suggest defining a task as a set of images from the same volume $\mathcal{V}$ sampled with the step size = $\left \lceil{\frac{|\mathcal{V}|}{K}}\right \rceil$. This ensures the balancedness of task sizes across datasets. %

\subsubsection{Weighted task sampling}
Some organs have different modalities and/or different zones (e.g. prostate can be split into peripheral and transitional zones), which we treat as separate datasets. One can see that this might lead to some organs dominating during task sampling (e.g. BRATS dataset alone is translated into 12 different source datasets). To counter that, we suggest using weighted sampling to give each organ a fair chance to get into the tasks set in each meta-epoch. For each organ with $z$ different modalities or zones, we set the sampling rate of each modality / zone to $\frac{1}{z}$ and normalize them. 

\subsubsection{Importance-aware task weighting}
We employ the original Reptile~\cite{nichol2018first} update rule as a baseline for our method:
\begin{equation}
\theta \leftarrow \theta +
  \beta  \frac{1}{L} \sum_{l=1}^{L}  {(\theta_l-\theta)},
\end{equation}
where $\theta_l$ is the weights vector of the local model, $\theta$ is the weights vector of the meta-model, $L$ is the number of tasks and $\beta$ is the meta learning rate. 
We compare the following settings:
\begin{itemize}
    \item Average weighting (AW): all updates have the same weights and $(\theta_l-\theta)$ are averaged across tasks.
    \item Inverse distance weighting (IDW): more weight is given to models closer to the meta-model. %
\end{itemize}
The weights for AW update are defined by the number of tasks sampled in each meta-epoch. With $L$ tasks sampled, the weights for each of tasks would be: $\frac{1}{L}$. The weighting for the inverse distance update is given by:
\begin{equation}
\label{eq:meta_update}
w_l = \frac{1}{\sum_{i=1}^{L}{(\theta_{l, i} - \theta_{i})^2 }},
\end{equation}
where $\theta_i$ is the $i$-th weight of the meta-model and $w_{l,i}$ is the $i$-th weight of the $l$-th task's model of the current meta-epoch. The weights are also normalized to sum up to 1 using $w_l = \frac{w_l}{\sum_j^L w_j}$. The update rule is therefore: 
\begin{equation}
\label{eq:meta_update2}
\theta \leftarrow \theta +
  \beta   \sum_{l=1}^{L} {w_l (\theta_l-\theta)},
\end{equation}

\subsection{Image Segmentation}
Binary cross entropy (BCE) is the common loss term for optimizing image segmentation models. However, our initial experiments demonstrated quite poor performance in a few-shot setting, so we experimented with other losses. We employ the weighted BCE loss in combination with the approximation of Intersection over Union (IoU) loss~\cite{Rahman2016} for segmentation.
\subsubsection{Weighted BCE Loss} Given an input image $\mathcal{I}$, predicted segmentation map $y$ and ground truth segmentation map $y'$, the weighted BCE loss is calculated as:
\begin{equation}
\label{eq:bceloss}
    BCE(y,y') = - (p_{\text{pos}} y \log (y') + (1 - y) \log (1 - y'))
\end{equation}

where $p_{\text{pos}}$ is the weight of positive samples (ratio of sum of object pixels to sum of background pixels). 
\subsubsection{IoU loss} For the IoU loss, we adopt the IoU approximation proposed by Rahman et al. in ~\cite{Rahman2016}, defined as:

\begin{equation}
\label{eq:iouloss}
\begin{aligned}
    \mathcal{X}(y,y') = \sum_{i \in \mathcal{P}} y_i * y'_i \\ %
    \mathcal{U}(y,y') = \sum_{i \in \mathcal{P}} (y_i +y'_i - y_i * y'_i) \\ %
    IoU(y,y') = \frac{\mathcal{X} + \epsilon}{\mathcal{U} + \epsilon},
    \end{aligned}
\end{equation}
where $\mathcal{P}$ is the set of all training images pixels, $y_i$ is the ground truth pixel value (0 for background, 1 for object), and $y'_i$ is the probability - the sigmoid output of the U-Net. We add $\epsilon$ to the formula for numerical stability.
Finally, we experiment with the combination of BCE and logarithmic DICE loss, using Equation~\ref{eq:ioubce} (see derivation in the supplementary material): 

\begin{dmath}
\label{eq:ioubce}
Q(L(y,y') = BCE(y,y')-\log\left(\frac{2 IoU(y,y')}{IoU(y,y') + 1}\right)
\end{dmath}

%% file: Chapters/results.tex
\section{Experiments}
\begin{table}[tbp]
    \centering
    \caption{Comparison of our proposed methods to related work in a few-shot setting, fine-tuned on 15 shots on 4 different organs. AW and IDW stand for Average weighting and Inverse distance weighting respectively.}
    \begin{tabu}{|c|c|c|c|}
    \hline
        \multirow{2}{*}{Update rule} & \multirow{2}{*}{Target organ} & \multicolumn{2}{c|}{IoU $\uparrow$} \\ \cline{3-4}
         &  &  Standard & Volume-based  \\\hline
        Supervised Learning & Cardiac & $58.78$ & $-$\\\hline
        Transfer Learning & Cardiac & $66.22$ & $-$\\\hline
        Dawoud et al. \cite{dawoud2020few} & Cardiac & $68.28$ & $67.7$\\\hline
        
        MetaMedSeg + IDW (Ours) & Cardiac & $67.49$ & $64.86$\\\hline
        MetaMedSeg + AW (Ours) & Cardiac & $67.83$ & $\textbf{68.33}$\\\hhline{|=|=|=|=|}
        
        Supervised Learning & Spleen & $38.81$ & $-$\\\hline
        Transfer Learning & Spleen & $51.18$ & $-$ \\\hline
        Dawoud et al. \cite{dawoud2020few} & Spleen & $49.29$ & $\textbf{58.34}$ \\\hline
        
        MetaMedSeg + IDW (Ours) & Spleen & $55.64$ & $50.50$\\\hline
        MetaMedSeg + AW (Ours) & Spleen & $55.98$ & $56.44$\\\hhline{|=|=|=|=|}
        
        Supervised Learning & Prostate Peripheral & $7.35$ & $-$\\\hline
        Transfer Learning & Prostate Peripheral & $10.87$ & $-$\\\hline
        Dawoud et al. \cite{dawoud2020few} & Prostate Peripheral & $15.99$ & $12.82$\\\hline
        
        MetaMedSeg + IDW (Ours) & Prostate Peripheral & $17.15$ & $13.89$\\\hline 
        MetaMedSeg + AW (Ours) & Prostate Peripheral & $16.17$ & $\textbf{22.69}$\\\hhline{|=|=|=|=|}
        
        Supervised Learning & Prostate Transitional & $38.64$ & $-$\\\hline
        Transfer Learning & Prostate Transitional & $41.09$ & $-$\\\hline
        Dawoud et al. \cite{dawoud2020few} & Prostate Transitional & $42.85$ & $46.28$\\\hline
        
        MetaMedSeg + IDW (Ours) & Prostate Transitional & $44.25$ & $44.72$\\\hline 
        MetaMedSeg + AW (Ours) & Prostate Transitional & $42.43$ & $\textbf{48.33}$\\\hline
        
    \end{tabu}
    \label{tab:comparison_few}
\end{table}

\begin{table}[htbp]
    \centering
    \caption{Comparison of our proposed methods to related work in full-data setting on 4 different organs. AW and IDW stand for Average weighting and Inverse distance weighting respectively.}
    \begin{tabu}{|c|c|c|c|}
    \hline
        \multirow{2}{*}{Update rule} & \multirow{2}{*}{Target organ} & \multicolumn{2}{c|}{IoU $\uparrow$} \\ \cline{3-4}
         &  &  Standard & Volume-based  \\\hline
         
        Supervised Learning & Cardiac & $90.26$ & $-$\\\hline
        Transfer Learning & Cardiac & $90.46$ & $-$\\\hline
        Dawoud et al. \cite{dawoud2020few} & Cardiac & $90.38$ & $92.47$ \\\hline
        
        MetaMedSeg + IDW (Ours) & Cardiac & $91.38$ & $94.51$\\\hline 
        MetaMedSeg + AW (Ours) & Cardiac & $91.08$ & $\textbf{95.55}$\\\hhline{|=|=|=|=|}
        
        Supervised Learning & Spleen & $86.74$ & $-$ \\\hline
        Transfer Learning & Spleen & $86.10$ & $-$\\\hline
        Dawoud et al. \cite{dawoud2020few} & Spleen & $89.65$ & $87.53$\\\hline
        
        MetaMedSeg + IDW (Ours) & Spleen & $89.96$ & $91.53$\\\hline
        MetaMedSeg + AW (Ours) & Spleen & $90.00$ & $\textbf{92.27}$
        \\\hhline{|=|=|=|=|}
        
        Supervised Learning & Prostate Peripheral & $ 39.06$ & $-$\\\hline
        Transfer Learning & Prostate Peripheral & $39.94$ & $-$\\\hline
        Dawoud et al. \cite{dawoud2020few} & Prostate Peripheral & $37.05$ & $41.58$\\\hline
        
        MetaMedSeg + IDW (Ours) & Prostate Peripheral & $47.58$ & $68.26$\\\hline
        MetaMedSeg + AW (Ours) & Prostate Peripheral & $46.20$ & $\textbf{70.90}$
        \\\hhline{|=|=|=|=|}
        
        Supervised Learning & Prostate Transitional & $68.04$ & $-$\\\hline
        Transfer Learning & Prostate Transitional  & $69.84$ & $-$\\\hline
        Dawoud et al. \cite{dawoud2020few} & Prostate Transitional  & $70.42$ & $71.55$\\\hline
        MetaMedSeg + IDW (Ours) & Prostate Transitional  & $68.98$ & $\textbf{79.95}$\\\hline  
        MetaMedSeg + AW (Ours) & Prostate Transitional  & $67.68$ & $78.84$\\\hline
        
    \end{tabu}
    \label{tab:comparison_all}
\end{table}

 We train and evaluate our approach on the medical decathlon dataset~\cite{simpson2019large}, which consists of 3D MRI and CT volumes of 9 organs, namely Brain (368 volumes), Hippocampus (260), Lung (25), Prostate (32), Cardiac (20), Pancreas (279), Colon (121), Hepatic Vessels (216), and Spleen (41). We use the U-Net~\cite{ronneberger2015u} architecture and train our model using IoU, BCE, and the combination of the two losses. \\
As a baseline, we implement the transfer learning approach, where we train with all available training data to obtain the initialization weights for fine-tuning. \\
To make sure that the results obtained in the experiments are not due to the specific k-shot selection we perform fine-tuning on 5 different random selections and test on the same test set of previously unseen data. The evaluation metric (IoU) is then averaged over the 5 runs.

\subsection{Experimental Setup}
The data is pre-processed by splitting different techniques (e.g. T2 and FLAIR) and different regions (e.g. edema and tumour) into separate datasets, which results in $24$ different datasets. For each dataset, we set a threshold indicating whether we consider an object present on the image or not based on the number of pixels and visual inspection of the results. The values of thresholds can be found in the supplementary material. All the images were resized to $256 \times 256$ resolution. The threshold was applied after resizing. 
We also apply volume normalization during slicing by subtracting the mean and dividing by the standard deviation of all the non-zero pixels of the whole volume. We fix the same conditions for all fine-tuning experiments: we train for 20 epochs using weight decay $3 \times 10^{-5}$, learning rate $\alpha=0.005$ with a step learning rate decay of $\gamma=0.7$ at every other step. For full-data training, we use a learning rate $0.001$ and weight decay $w=3 \times 10^{-5}$. 
For meta-training we train for $100$ epochs, sampling $5$ tasks with $15$ shots and $1$ image per shot at each meta-epoch. We used a learning rate of $\alpha=0.01$ for local and meta-model with weight decay of $w=0.003$ and the same learning rate decay described above. For transfer learning, we train on all datasets excluding cardiac, prostate, and spleen for $20$ epochs. %

To compare with \cite{dawoud2020few}, we use the same hyperparameters as in the original paper for meta-training. It should be noted that, following the protocol from \cite{dawoud2020few}, we go sequentially over all the available source datasets rather than sampling them. Also, following the same protocol, for each task, we sample 2 additional tasks for 2 additional losses from different source datasets. 
For fine-tuning we also follow \cite{dawoud2020few}, but train for 40 epochs instead of 20 to get better performance and also use IoU loss instead of BCE.

\subsection{Comparison to Related Work}
\begin{figure}[htbp]
    \centering
    \includegraphics[width=0.9\textwidth]{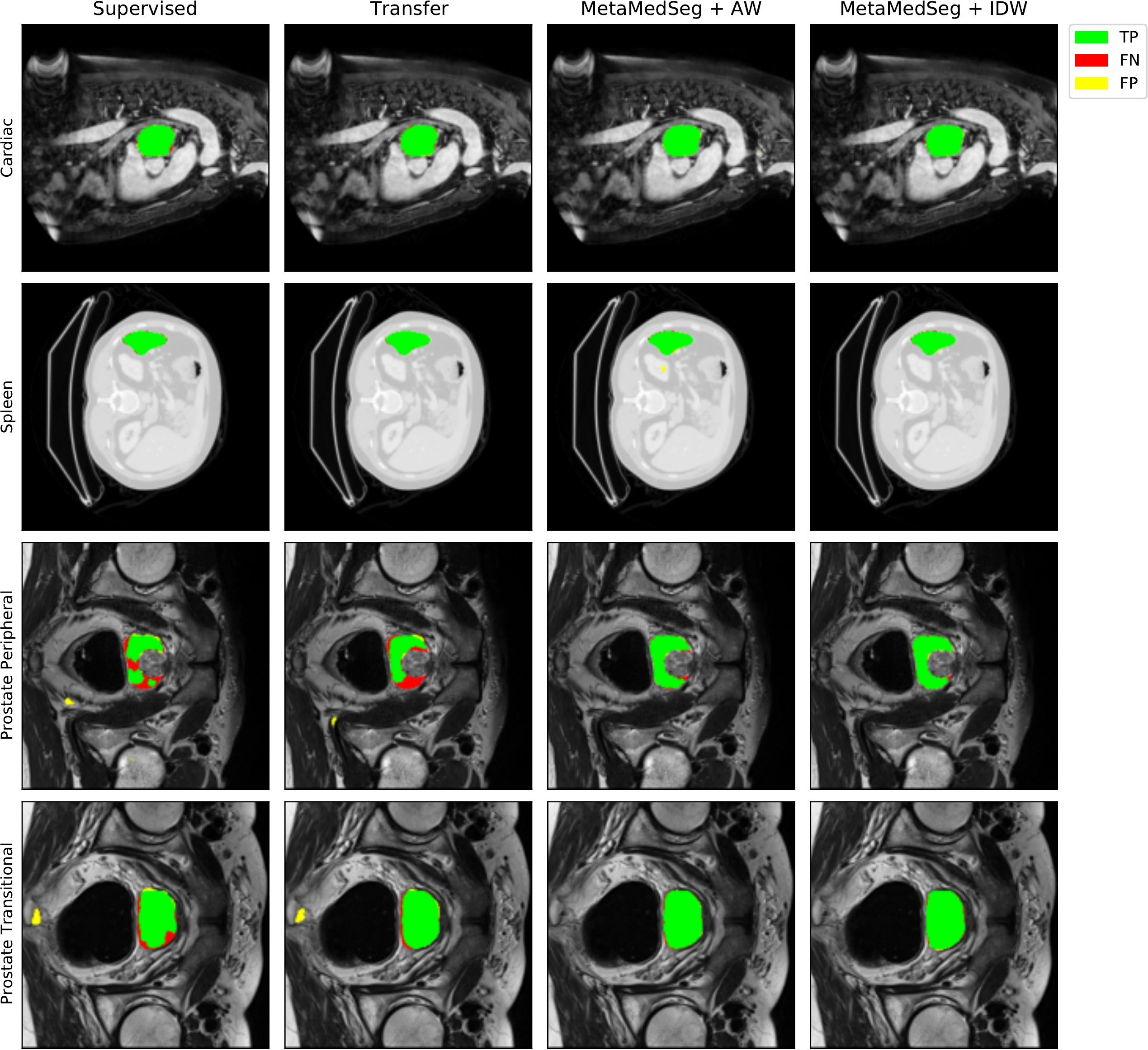}
    \caption{A comparison of different segmentation baselines with our method for four different target organs in the full-data setting.}
    \label{fig:qualitative_results}
\end{figure}

\Cref{tab:comparison_few} shows the comparison of our work to the baseline methods in few-shot setting for $15$ shots on four different organs. The performance of the models fine-tuned on the whole support set is shown in \cref{tab:comparison_all}. Some of examples of segmentation results are shown in ~\cref{fig:qualitative_results}.

\subsection{Discussion}
\begin{figure}[htbp]
    \centering
    \includegraphics[width=0.74\textwidth]{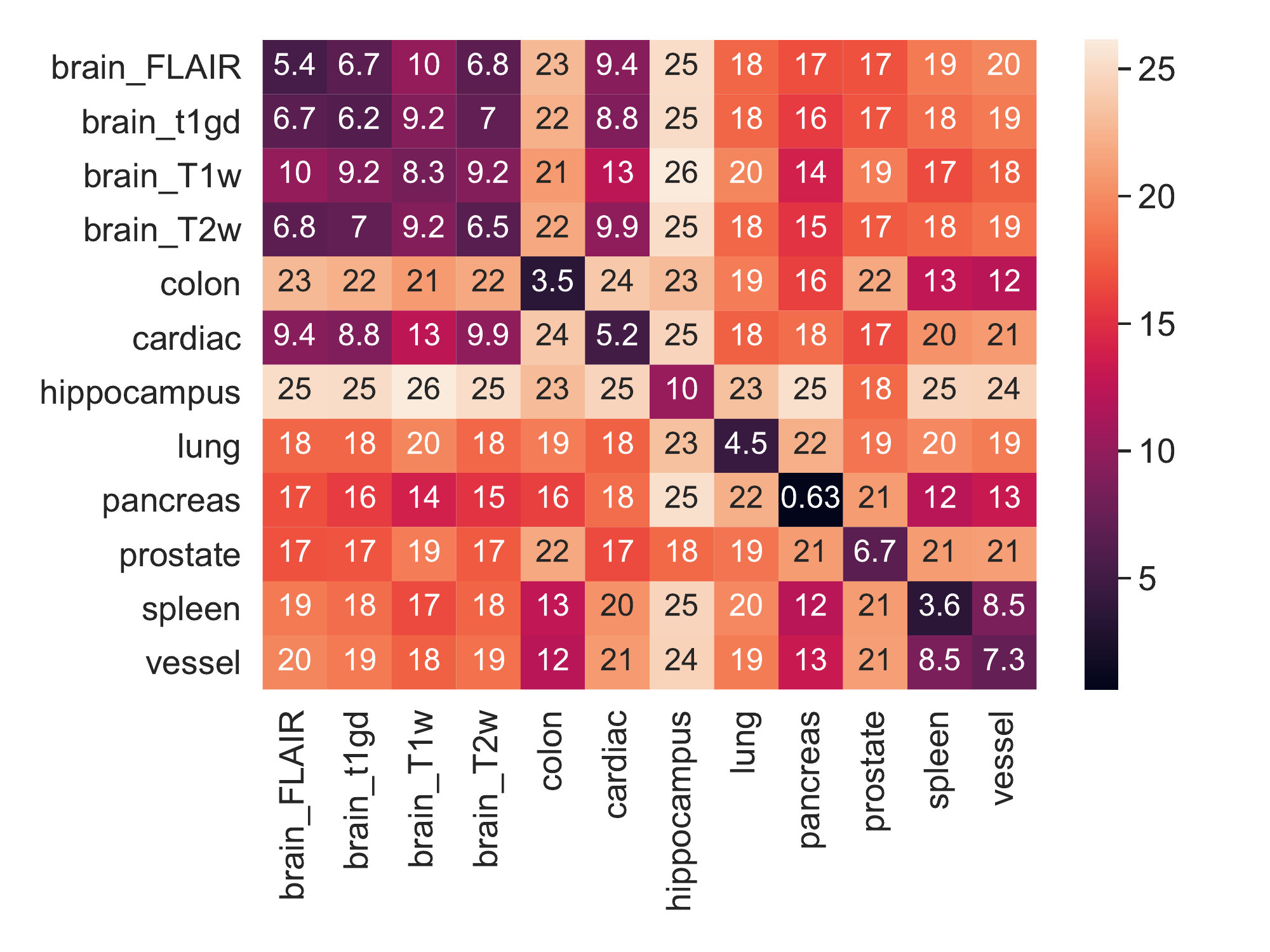}
    \caption{Heatmap of the average distances between pairs of images from different organs}
    \label{fig:adjmat}
\end{figure}
The results of our experiments show that volume-based task design has the most effect on the segmentation performance, especially in the full-data setting. The effect of our proposed update rule is minimal in some cases, but visible in other organs such as Prostate Transitional. This could be due to the nature of data distribution and the shapes of organs. When organ shapes have a high diversity (e.g. prostate compared to other organs), the outlier shapes could benefit from the proposed inverse distance update rule which gives less weight to gradients further from average. We hypothesize that, when using volumetric task definition, all organs have the chance to contribute to the final model in a more balanced setting. Therefore, when the weighted update rule is combined with volumetric tasks, the performance decreases. Another reason could be that the weighting of the updates helps to bias the model towards tasks that are more similar to the target. To understand the effect of image diversity, we create an adjacency matrix of the average Euclidean distances between pairs of randomly selected volumes. The results in ~\cref{fig:adjmat} show that in our target organs the following have the lowest to highest distance: cardiac, spleen, prostate. The effect of this distance is visible in the final segmentation performance of each organ.

\subsection{Ablation Study}
We show the effect of the proposed update rule, different segmentation losses, and the volume-based task definition used in this work in ~\cref{tab:ablation}. The models were meta-trained using five different losses including Focal Tversky loss~\cite{abraham2019novel} and Dice loss~\cite{sudre2017dice}, and then fine-tuned using IoU loss. The best performance is achieved by using weighted BCE loss for meta-training and IoU loss for fine-tuning.

\begin{table}[htbp]
    \centering
    \caption{Ablation study of our method with different losses in few-shot setting for cardiac segmentation. AW and IDW stand for Average weighting and Inverse distance weighting respectively.}
    \begin{tabular}{|c|c|c|c|c|}
    \hline
        \multirow{2}{*}{Update rule} & \multirow{2}{*}{Segmentation loss} & \multicolumn{2}{c|}{IoU $\uparrow$} \\ \cline{3-4}
         &  &  Standard & Volume-based  \\\hline
        AW & IoU & $67.82$ & $68.13$\\\hline
        IDW & IoU & $67.42$ & $64.86$\\\hline
        
        AW & Tversky Focal loss~\cite{abraham2019novel} & $65.90$ & $65.72$\\\hline
        IDW & Tversky Focal loss~\cite{abraham2019novel} & $63.71$ & $62.78$\\\hline
        
        AW & Dice loss~\cite{sudre2017dice} & $65.87$ & $66.03$\\\hline
        IDW & Dice loss~\cite{sudre2017dice} & $62.58$ & $62.73$\\\hline

        AW & BCE & $\textbf{67.83}$ & $\textbf{68.33}$\\\hline
        IDW & BCE & $65.09$ & $64.29$\\\hline
        
        AW & BCE + IoU & $66.85$ & $67.30$\\\hline
        IDW & BCE + IoU & $66.98$ & $64.71$ \\\hline
    \end{tabular}
    \label{tab:ablation}
\end{table}

%% file: Chapters/conclusion.tex
\section{Conclusion}
We presented a novel way of task definition for few-shot learning for volume-based 2D data and an update rule based on the importance of tasks in the meta-training step. Our method is evaluated on four different organ types with the least amount of data, namely cardiac, spleen, prostate peripheral and prostate transitional. Our approach is not only applicable to organ segmentation but also to tumour segmentation or other types of densities. The results show that our proposed volumetric task definition improves the segmentation performance significantly in all organs. The proposed update rules provide considerable improvement in terms of IoU. Both proposed approaches (volumetric tasks and weighted update rule) could be useful in different scenarios. While the volumetric task definition proves to be advantageous in all scenarios, it is more beneficial to use the weighted update rule when the data distribution of the target class is different from the source, for example in cases such as segmentation of new diseases where also the amount of labelled data is limited.

\section*{Acknowledgement}
We gratefully acknowledge the Munich Center for Machine Learning (MCML) with funding from the Bundesministerium für Bildung und Forschung (BMBF) under the project 01IS18036B. We are also thankful to Deutsche Forschungsgemeinschaft (DFG) for supporting this research work, under project 381855581.